\documentclass[11pt,prd,letterpaper,amsmath,amssymb,final,nofootinbib, 
,unsortedaddress,superscriptaddress]{revtex4-1}

\usepackage{amsmath}
\usepackage{graphicx}
\usepackage{multirow}
\usepackage{dcolumn}
\usepackage{bm}

\usepackage[mathlines]{lineno}

\begin{document}

\preprint{AIP/123-QED}

\title[]{The Physics and Nuclear Nonproliferation Goals of WATCHMAN: A WAter CHerenkov Monitor for ANtineutrinos}

\author{M.~Askins}
\affiliation{Physics Department, University of California, Davis CA 95616, USA}

\author{M.~Bergevin}
\affiliation{Physics Department, University of California, Davis CA 95616, USA}

\author{A.~Bernstein\footnote{Corresponding Author, email: bernstein3@llnl.gov}}
\affiliation{Lawrence Livermore National Laboratory, Livermore, CA 94550, USA}

\author{S.~Dazeley}
\affiliation{Lawrence Livermore National Laboratory, Livermore, CA 94550, USA}

\author{S.~T.~Dye}
\affiliation{Department of Natural Sciences, Hawaii Pacific University, Kaneohe, Hawaii 96744, USA}

\author{ T. Handler}
\affiliation{Department of Physics and Astronomy, University of  Tennessee,Knoxville TN 37996}

\author{ A. Hatzikoutelis}
\affiliation{Department of Physics and Astronomy, University of  Tennessee,Knoxville TN 37996}

\author{D. Hellfeld}
\affiliation{Department of Nuclear Engineering, Texas A\&M University, College Station,  TX, 77843}

\author{P. Jaffke}
\affiliation{Department of Physics, Virginia Polytechnic Institute and State University, Blacksburg, VA 24061, USA}

\author{ Y. Kamyshkov}
\affiliation{Department of Physics and Astronomy, University of  Tennessee,Knoxville TN 37996}

\author{B.~J.~Land}
\affiliation{Lawrence Berkeley National Laboratory, Berkeley, CA 94720, USA}
\affiliation{Department of Physics, University of California, Berkeley, CA 94720, USA}

\author{J.~G.~Learned}
\affiliation{Department of Physics and Astronomy, University of Hawaii at Manoa, Honolulu, HI 96922 USA}

\author{P.~Marleau}
\affiliation{Sandia National Laboratories, Livermore, CA 94550, USA}

\author{C. Mauger}
\affiliation{Los Alamos National Laboratory, Los Alamos, NM 87544m, USA }

\author{G.~D.~Orebi Gann}
\affiliation{Lawrence Berkeley National Laboratory, Berkeley, CA 94720, USA}
\affiliation{Department of Physics, University of California, Berkeley, CA 94720, USA}

\author{C. Roecker}
\affiliation{Department of Nuclear Engineering, University of California, Berkeley, CA 94720, USA}
\affiliation{Lawrence Livermore National Laboratory, Livermore, CA 94550, USA}
\affiliation{Sandia National Laboratories, Livermore, CA 94550, USA}

\author{S.~D.~Rountree}
\affiliation{Department of Physics, Virginia Polytechnic Institute and State University, Blacksburg, VA 24061, USA}

\author{T.~M.~Shokair}
\affiliation{Department of Nuclear Engineering, University of California, Berkeley, CA 94720, USA}

\author{M.~B.~Smy}
\affiliation{Department of Physics and Astronomy,
University of California, Irvine, CA 92697, USA}

\author{R.~Svoboda}
\affiliation{Physics Department, University of California, Davis CA 95616, USA}

\author{M.~Sweany}
\affiliation{Sandia National Laboratories, Livermore, CA 94550, USA}

\author{M.~R.~Vagins}
\affiliation{Department of Physics and Astronomy,
University of California, Irvine, CA 92697, USA}

\author{K.~A.~van~Bibber}
\affiliation{Department of Nuclear Engineering, University of California, Berkeley, CA 94720, USA}

\author{R.~B.~Vogelaar}
\affiliation{Department of Physics, Virginia Polytechnic Institute and State University, Blacksburg, VA 24061, USA}

\author{M.~J.~Wetstein}
\affiliation{Enrico Fermi Institute, University of Chicago, Chicago, IL 60637, USA}

\author{M.~Yeh}
\affiliation{Brookhaven National Laboratory, Upton, NY 11973, USA}


\date{\today}

\begin{abstract}
This article describes the physics and nonproliferation goals of  WATCHMAN, the WAter Cherenkov Monitor for ANtineutrinos. The baseline WATCHMAN design is a kiloton scale gadolinium-doped (Gd) light water Cherenkov detector, placed 13 kilometers from a civil nuclear reactor in the United States. In its first deployment phase, WATCHMAN will be used to remotely detect a change in the operational status of the reactor, providing a first-ever demonstration of the potential of large Gd-doped water detectors for remote reactor monitoring for future international nuclear nonproliferation applications.  A demonstration of remote monitoring of a reactor has been called for in the U.S. National Nuclear Security Adminstration's (NNSA) Strategic  Plan \cite{stratplan}. 

During its first phase, the detector will also provide a critical large-scale test of the ability to tag neutrons and thus distinguish electron-flavor neutrinos and antineutrinos, using gadolinium-doped water. This would make WATCHMAN the  world's only detector capable of providing both direction and flavor identification of supernova neutrinos. It would also be the world's third largest supernova detector,  and the largest in the western hemisphere.

In the first and subsequent phases, the detector will also be a flexible test-bed for a range of large doped water Cherenkov detector technologies, including high quantum efficiency  photomultipliers, Water-Based Liquid Scintillator (WbLS), picosecond light sensors, advanced image recognition methods, and other enhancements.   A separate white paper describing Theia, a WbLS-based Advanced Scintillator Concept Detector (ASDC) has been published, with a long-term goal of fielding a 50-100 kton WBLS detector situated at the Long Baseline Neutrino Facility (LBNF). As described in that paper, WATCHMAN is essential to the demonstration of the advanced technologies that will be incorporated into the ASDC \cite{Alonso:2014fwf}. The physics enabled by this technology development is described in detail in the ASDC paper, including mass hierarchy measurements, measurement of CP violation in the neutrino sector, and neutrino-less double beta decay. 

Finally, if a compact accelerator is placed nearby, WATCHMAN will enable a highly sensitive search for sterile neutrinos with few-meter oscillation wavelengths, and could perform a search for Non-Standard neutrino Interactions (NSI). The sensitivity in both cases would increase if a WbLS or oil-based liquid scintillator (LS) target is used. The compact accelerator is now being developed by the ISoDAR collaboration \cite{ISODAR}.

This white paper describes the WATCHMAN conceptual design, including the results of a detailed simulation of the signal and backgrounds, discusses our current Preferred and Backup Alternative deployment sites, and reviews the current status of the WATCHMAN nonproliferation and physics programs.

\end{abstract}

\pacs{Valid PACS appear here}

\maketitle


\section{Introduction}

Gadolinium-doped water has long been proposed as a  medium for large scale neutrino and antineutrino detectors \cite{doi:10.1080/08929880108426496},\cite{Beacom:2003nk}. Gadolinium doping is recognized as essential for detection of low energy antineutrinos with light water Cherenkov detectors. The technique has been proposed by the LBNE collaboration, because it would permit sensitive studies of supernovae,  cosmological/relic neutrinos and possibly proton decay  \cite{2012arXiv1204.2295T}.  Among its other attractions, WATCHMAN provides a timely and highly subsidized opportunity to demonstrate the gadolinium option for LBNE.


The  gadolinium dopant ensures efficient  detection of the neutron  created in charged-current antineutrino-proton scattering, known as inverse beta decay:
\begin{equation}
\bar{\nu}+p \rightarrow e^{+} +n.
\end{equation}

Here $\bar{\nu}$ is the incident antineutrino, $p$ the target proton, and $e^{+}$ and $n$ the final state reaction products, a positron and a neutron. 

Compared with a pure water Cherenkov detector, gadolinium doping provides the important advantage of an approximate six-fold increase in the efficiency for detection of the final state neutron. This benefit derives from two useful features of the gadolinium nucleus: its high thermal neutron capture cross-section, and the relatively high energy gamma-ray emissions arising from the post-capture excited Gd nucleus. Even at $0.1\%$ doping (as measured by the weight of the Gd element), gadolinium captures about $85\%$ of all neutrons arising from the inverse beta decay process. The $\sim \!\! 8 \, MeV$ gamma-cascade following capture generates a strong Cherenkov pulse that is well above most backgrounds. This time-correlated signal  associates the neutron with the energy deposition arising from the positron which precedes it by a few tens of microseconds, with the mean time interval determined by the thermalization and capture time of the neutron in the Gd-doped water. 

In the last ten years, significant progress has been made in demonstrating the potential of gadolinium-doped water as an antineutrino (and neutron) detection medium.  Important milestones include the  demonstration of high efficiency detection of time-correlated neutrons in a ton-scale Gd-doped water Cherenkov detector \cite{Dazeley:2008xk}, and detection of gamma-rays arising from a Gd captures in a small enclosed cell of Gd-doped water placed near the center of the SuperKamiokande 50,000 ton pure water detector \cite{Watanabe2009320}. Very recently, the EGADS experiment in Japan has demonstrated on a large scale (200 tons) the long-term compatibility of Gd-doped water with typical detector materials such as stainless steel and Photomultiplier Tubes (PMT’s), and the ability to recirculate and continuously purify the water via selective extraction and reintroduction of the gadolinium, using a so-called 'molecular band-pass' filtration system \cite{EGADS}.

While the KamLAND  liquid scintillator detector has shown sensitivity to reactor antineutrinos at ~200-400 km standoff distances \cite{PhysRevLett.90.021802}, direct sensitivity to reactor antineutrinos has not yet been demonstrated with a gadolinium-doped water detector. The purpose of the WATCHMAN project is to perform such a demonstration, with important immediate and long-term benefits for  fundamental physics and for international nonproliferation.  

To this end, the WATCHMAN collaboration will deploy a gadolinium-doped water detector a few kilometers from a nuclear reactor.  The current preferred deployment site is the Morton Salt mine in Fairport Ohio, about 13 kilometers  from the single core Perry Nuclear Power Plant. The mine operators have approved use of this  deployment location, the former deployment site for the Irvine-Michigan-Brookhaven (IMB) detector \cite{BeckerSzendy:1992hr}.  


The demonstration will represent a major step towards a true long range reactor monitoring capability for nonproliferation purposes. WATCHMAN will also have a strong fundamental neutrino physics program. It will be the United States' only supernova watch detector and will have the unique capability of flavor identification and neutrino direction in a single detector. It would be the third largest SN detector in the world. In addition, the same detector may be used to search for sterile neutrino oscillations and non-standard neutrino interactions using a nearby compact cyclotron to create an intense neutrino flux. 


 The WATCHMAN effort also overlaps closely with plans and technology needs in the fundamental science community, for very large scale detectors to study astrophysical neutrinos,  neutrino oscillations, and other physical phenomena such as proton decay and the neutrino mass hierarchy. There is significant overlap between the membership of the WATCHMAN collaboration and  that of the Advanced Scintillator Detector Concept (ASDC) \cite{Alonso:2014fwf}. The ASDC initiative is engaged in planning and design studies for a multi-hundred-ton WBLS detector to be built  in the United States.  Because of this overlap, WATCHMAN can test  technologies that are essential to the success of ASDC and similar future detectors, in whatever country they are ultimately constructed. Examples include the  basic WATCHMAN technology of gadolinium doping and recirculation of gadolinium-doped water, as well as testing of water-based liquid scintillator, advanced photomultiplier tubes such as the Large Area Avalanche Photodetector (LAPPD) \cite{LAPPD}, wavelength-shifting plates, wireless readout, and other innovations. 

In this paper, we introduce the baseline design, site, and detector performance characteristics.  We then describe how the nonproliferation goals will be met with the baseline design. Next we provide analyses of sensitivity for each of the physics goals described above. We conclude with an overview of progress to date, the likely timeline of the experiment, and its relation to other large scale detectors being contemplated in the U.S. and worldwide.

\section{Site Selection}
\label{sec:sites}

The following detection criteria have been defined regarding the nonproliferation goal of demonstrating remote reactor monitoring capability with Gd-H2O technology. These criteria helped determine the depth and standoff distance for the site:

\begin{enumerate}
\item A signal to background ratio sufficient for $99.7\%$ C.L. ($3 \sigma$) sensitivity to the presence of a reactor operating at full power, based on no more than roughly 30 days each of reactor-off and reactor-on data. This implies sensitivity to a single reactor on-off transition, which we consider most attractive from a nonproliferation standpoint. 
\item A detector standoff distance no closer than 500 meters from the reactor. 
\item  The detector is meant to demonstrate that low event rate detection is a viable tool for the nonproliferation community.  Therefore, we imposed a requirement that the total event rate be below a subjective threshold for rare event detection -  here defined as no greater than roughly 10 reactor antineutrino events per day. A higher event rate would mean that either our choice of baseline was too close, or our detector mass larger than necessary to allow a convincing demonstration of standoff detection with relatively low statistics signal. 

\end{enumerate}

Other  design and  cost criteria also influence the choice of site. These are: 
\begin{enumerate}
\item A detector target mass of approximately one thousand tons.
\item A detector target composition of H2O + Gd-salt, with about $0.1\%$ wt percent Gd concentration.
\item An initial detector design that includes a water recirculation system capable of selective capture and reintroduction of gadolinium (EGADS-like system). 

\end{enumerate}

Using these criteria, we performed an extensive survey of domestic reactor facilities and their surroundings, considering detector deployments at both underground and underwater sites. Excavation costs are a key concern for underground sites. To keep excavation costs low, we looked for already excavated sites near a reactor. The Morton Salt mine in Ohio was the sole site in the US that met this criterion. It was also found to be compatible with the other detection criteria defined above. 

In addition to the Ohio site, we identified a greenfield excavation deployment at the Advanced Test Reactor \cite{ATR}, near Idaho Falls in the state of Idaho. We found the geology to be compatible with a relatively shallow 300 mwe overburden tunnel, and a reactor standoff distance of approximately 1 kilometer. However, at this shallow depth, the higher and less well understood backgrounds would make the measurement more challenging and riskier.  The ATR site remains as a backup option in our planning process.

\subsection{The Morton Salt Mine and the Perry Nuclear Power Plant}

The deployment site for WATCHMAN is the Morton Salt Mine in Fairport, Ohio, about 13 kilometers from the single core Perry Nuclear Power Plant.  Figure~1 
shows the location of the power plant and mine along the shores of Lake Erie. Morton Salt has given our collaboration permission  to deploy the WATCHMAN detector at their facility on a cost neutral basis.

\begin{figure}[ht]
\begin{center}
\includegraphics[width=0.8\textwidth]{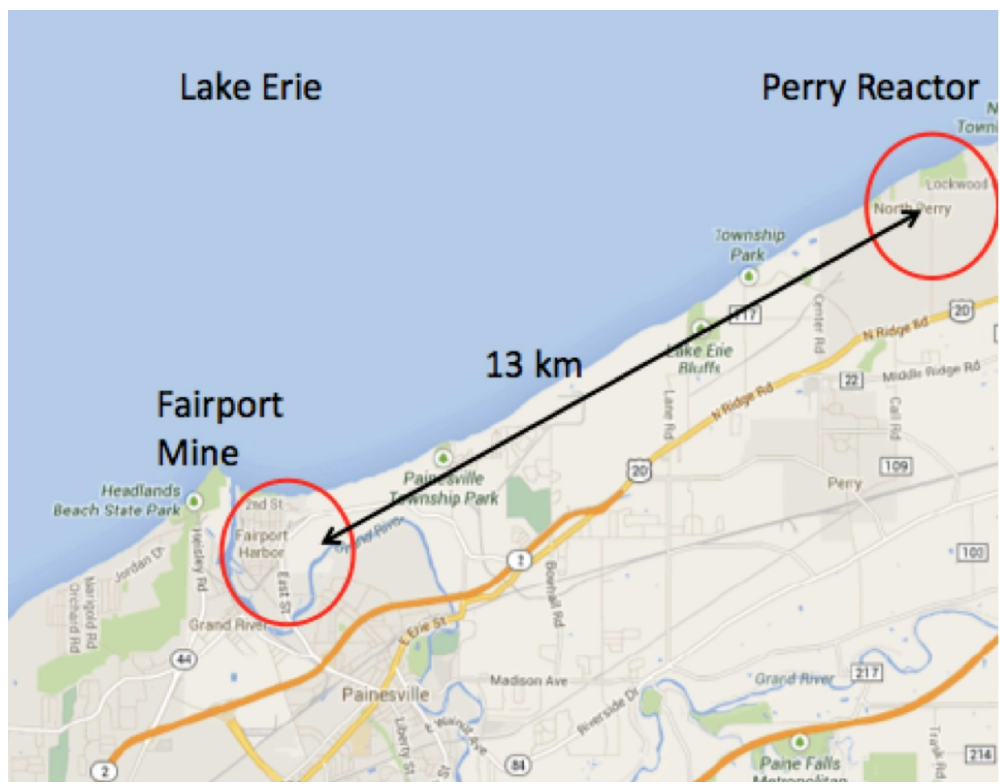}
\caption{The Morton Fairport Harbor Mine is 13 km from the Perry Nuclear Reactor. The site is convenient to reach---roughly 25 miles east of Cleveland, Ohio---and has a long history of supporting DOE science experiments.}
\end{center}
\label{fig:FairPortMap}
\end{figure}

Due to the depth and the proximity to the Perry reactor, this site is especially well suited to a reactor monitoring experiment. In addition, the Morton company has a long history of hosting DOE experiments. From 1980-1991, the underground laboratory site housed the IMB proton decay experiment. As a result, the site has an existing cavern that allows the project to avoid an otherwise costly greenfield excavation.  The layout of the cavern and laboratory are shown in Figure 2. 
The  cavern, and the remaining IMB laboratory infrastructure permit a quick and inexpensive refurbishment of the site with little risk of delay. Access is via a personnel lift roughly 100 meters from the lab's entrance; a separate, nearby ore lift serves as a secondary egress.  It should be noted that IMB was a 10,000 ton undoped water Cherenkov detector with a threshold of about 20 MeV. This high threshold and the lack of a gadolinium dopant made the detector incapable of detecting reactor antineutrinos.  However, the excavated IMB cavern is an excellent space for the WATCHMAN deployment, significantly reducing the project cost. Further, the similarity of the technologies gives strong confidence that it is technically possible to deploy our smaller 1000 ton water Cherenkov detector at this site, greatly reducing risk compared to other options. 

The Perry boiling water nuclear reactor is a single core commercial plant with a thermal power rating of 3,758 MegaWatts (MWt). It is one of the most powerful of this reactor type in the country. As one of the newest reactors built in the U.S., it is licensed to operate until March 18, 2026 and has applied for a 20 year lifetime extension.  Either date is well beyond the period needed for all phases of WATCHMAN. In addition, the Morton Salt Mine, which supplies road salt to the Great Lakes area, has an essentially unlimited operational future since the salt deposit extends for many miles under the lake.

\begin{figure}[ht]
\begin{center}
\includegraphics[width=1.0\textwidth]{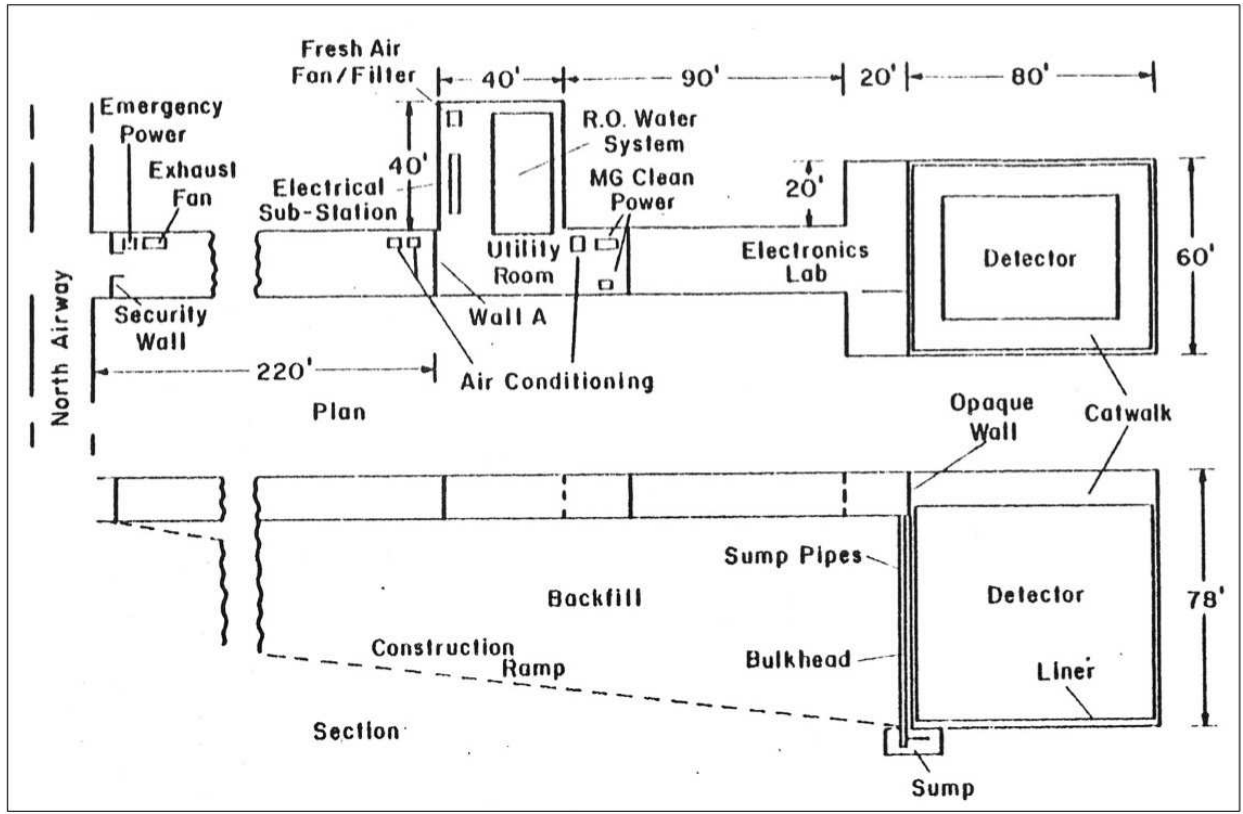}
\caption{The existing underground lab layout at the Morton salt mine in Ohio. The WATCHMAN deployment will reuse much of this space.}
\label{IMBlab}
\end{center}
\end{figure}

The Perry option also provides a demonstration of sensitivity at greater standoff – about 13 km compared with $\sim1-2$ km for the research reactors we examined - albeit with a higher power core.

\subsection{Reactor duty cycle}
The Perry plant has a typical outage cycle of about 40-50 days. Its single core is an advantage for our nonproliferation goal, since the contrast with background is increased compared to sites with more than one core.  

\subsection{Site approvals for deployment}
In July 2013, members of the WATCHMAN collaboration met with Morton plant personnel at the mine.  An inspection of the existing cavern showed it to be intact and suitable to accommodate the WATCHMAN detector.  Mine management has indicated a willingness to host the WATCHMAN detector on a cost-neutral basis.

\section{Nonproliferation Demonstration}

The main purpose for the initial phase of WATCHMAN is to demonstrate high efficiency detection of reactor antineutrinos in a kiloton-scale gadolinium-doped water (Gd-H2O) Cherenkov detector.  Gd-H2O is one of a few media, and possibly the only medium for which construction of a 100-1000 kiloton scale detector is achievable on grounds of cost and environmental impact.  The WATCHMAN deployment will demonstrate many of the physics and engineering features of the larger detectors, and will serve as a guide for their design and deployment. At the kiloton scale, the detector would be capable of excluding the existence of an operating 10 MWt reactor with high confidence in a $\sim\!25$ kilometer radius. This may itself have some limited utility in specific states, in which the international community might seek to non-intrusively and cooperatively confirm the non-operation of either known or unknown reactors within this radius. The potential utility becomes greater if the radius can be extended to several hundred kilometers, enabling possible cross-border detection. This requires large-scale detectors such as those now being contemplated for deployment to pursue fundamental physics goals. The most well-developed current example is the proposed Hyper-Kamiokande experiment \cite{HyperKLOI}.

\section{Baseline Design}
\label{sec:baseline}

The baseline WATCHMAN design uses a Gd-H2O target surrounded by a connected Gd-H2O veto region. The design is constrained by the size of the cavern space available at the preferred location at Fairport salt mine in Ohio. The detector vessel is a large self-supporting 300 series stainless steel tank which must fit in an area of 24.4 meters by 18.3 meters by 23.8 meters. The diameter and height of this cylindrical tank are 15.8 meters, with a total water volume of 3540 tons, accounting for displaced water volumes.  The detector is divided into two physical regions by a PMT support structure inside the tank that separates the inner target region from the outer veto region.   The target region is filled with 1810 tons of Gd-doped water, and the outer region is filled with 1730 tons of Gd-doped water. Inward-facing target region PMTs line the inside of the support structure, and opposite-facing PMTs point into the veto region from the same structure.  For our baseline design, the assumed fractional PMT coverage for the surfaces of the target and veto regions are $40\%$ and $4\%$ respectively. The effects of variations in the target PMT coverage are being examined by the collaboration.  

The face of the target PMTs is at a diameter of 12.8 meters. Within the region enclosed by the target PMTs, a smaller virtual fiducial volume is defined, based on the positions of the particle interactions as reconstructed from the variation of arrival times of Cherenkov light at the PMTs.   For a 1000 ton fiducial volume Gd-water detector, the diameter and height of this fiducial region is 10.82 meters. The region outside of the fiducial but inside the PMTs is referred to as the buffer region. The buffer serves to reduce the rate of gamma-rays and neutrons in the fiducial region arising from the PMTs and external radiation.

\begin{figure}
\begin{center}
\includegraphics[width=0.8\textwidth]{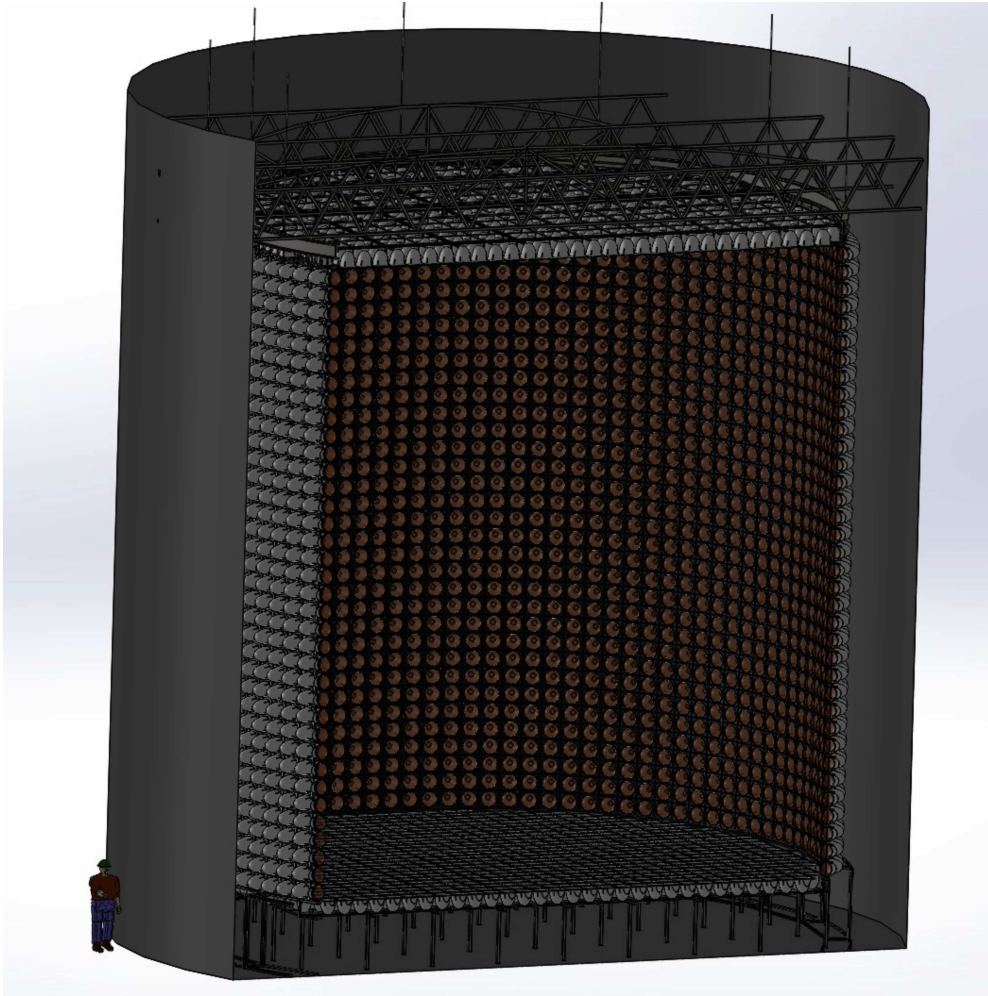}
\caption{ A cutaway view of the WATCHMAN detector (right) showing the gadolinium-doped water target region, the inward-facing (target) and outward facing (veto) PMTs, and cosmic ray muon veto region.}
\end{center}
\label{fig:detector}
\end{figure}

Figure~\ref{fig:detector} shows a cutaway drawing of the WATCHMAN detector. The PMT€™s are read out via fast ($<$1 ns resolution) digitizing electronics. The arrival time and intensity of the flashes can be reconstructed to give both the antineutrino interaction vertex and energy.

The most economical and practical method for this deployment is to have the tank fabricated from panels that are bolted together to form the cylinder.  These bolted tanks reduce the contamination present in welded tanks and are faster and less labor intensive to construct. Sealing material will used between the panels to avoid the need for internal tank liners, with the material experimentally screened to ensure no adverse reactions with the Gd-doped water or WbLS. The tank design is also flexible enough to allow a range of options for the location of fluid interconnects and access ports.

Instrumenting the veto and target regions requires large numbers of submersible PMTs. Among other choices, Hamamatsu $12^{\verb+"+}$  High Quantum Efficiency (HQE) PMTs are being considered as the light sensor for this application.  Based on preliminary simulations,  4328 such PMTs would be required to instrument the fiducial volume and approximately 482 for the veto region.   2880 inward-facing PMTs will cover the cylinder walls and an additional 724 PMTs will line the top and bottom surfaces of the cylinder. The remaining 482 PMTs will face the outer walls, top and bottom of the tank forming a veto region.  All the PMT assemblies will be mounted to modular framing and/or tension cables that provide positioning and support.  Based on previous large volume detector designs, specifically the LBNE project, our PMTs will be mounted in injection molded plastic housings that allow each PMT assembly to be secured from support structure in precise locations.

Table I shows the dimensions of the detector.

\begin{table}[htbp]
    \begin{tabular}{|c|c|c|c|}
\hline
    \textbf{Region}  &  \textbf{Mass (tons)} &  \textbf{Height (m)} &  \textbf{Diam./Thickness (m)} \\
    \textbf{Fiducial  (Gd-H2O)} & 1000  & 10.8  & 10.8  (diameter) \\
    \textbf{Buffer (Gd-H2O) } & 1049  & 12.8  & 1.0  (thickness) \\
    \textbf{Veto region (H2O)} & 1070  & 15.8  & 1.0  (thickness) \\
\hline  
\end{tabular}
\caption{The dimensions of the WATCHMAN baseline detector.}
\label{tab:dims}
\end{table}

\subsection{WATCHMAN operational phases}

The initial deployment of WATCHMAN will employ a Gd-H2O target. The WATCHMAN program will accommodate a second  phase, in which the target medium is changed to light WbLS, with a relatively low concentration of scintillator, and possibly a third phase using pure scintillator. The second phase in particular would allow demonstration of many of the key technologies to be used in the ASDC/Theia detector, including LAPPDs, high quantum efficiency photomultiplier tubes, and the light WbLS itself. The second and third phases will also accommodate a wider range of physics goals, primarily by increasing the sensitivity to sterile neutrinos in conjunction with the IsoDAR beam.


%
%
%

\section{Simulations of Reactor Signal and Backgrounds}
\label{sec:sbsims}

In this section we describe the reactor inverse beta decay signal and expected backgrounds in the baseline WATCHMAN design, and present estimates of the signal detection efficiency and background rejection capabilities of the detector.  These estimates are based  on a detailed simulation of the detector using the GEANT-based Reactor Monitoring SIMulation (RMSIM) package, maintained by UC Davis. The vertex reconstruction of events was made with the BONSAI package maintained at UC Irvine.  Using the same simulation package, we treat background and signal for other physics processes elsewhere in the paper. 

As described in the introduction, reactor antineutrinos interact with the quasi-free protons in water to produce a positron and neutron through the Inverse Beta Decay (IBD) process. The two final state particles generate Cherenkov light, which is detected by the PMTs surrounding the Gd-H2O target volume. The Cherenkov flash from the positron is proportional to the energy of the incident antineutrino. It induces single photoelectron pulses in a few dozen PMTs.  Each pulse has a width of a few tens of $ns$ defined by the specific PMT properties and optical dispersion and scattering in the water. This  prompt signal is closely followed by a delayed Cherenkov signal with a similar  time width and a comparable number of struck PMTs. The delayed signal arises from a cascade of multiple $MeV$-scale gamma-rays (summing to  $\sim8~MeV$ ) that are produced upon de-excitation of the gadolinium nucleus,  following its excitation by capture of the IBD-generated neutron. The time interval between the prompt positron light and the delayed neutron light is about 20-50 $\mu s$ depending on the gadolinium loading. Compared with backgrounds, the IBD event pairs are separated by a relatively short time interval. This temporally close event pair is often referred to as a correlated or coincident signal.   The pair of events are also relatively close spatially, and have relatively high energies compared to most backgrounds. As our simulations show, this combination of properties can be used to robustly suppress backgrounds and isolate a sample of antineutrino events with good efficiency.  

In Gd-H2O detectors, backgrounds for the reactor antineutrino signal fall into four main categories: 

\begin{enumerate}
\item pairs of accidental coincidences;
\item pairs of muogenic fast neutrons;
\item beta-n decays from  long-lived muogenic radionuclide on oxygen in the water; and, 
\item Other reactors
\end{enumerate} 

The energies of geological beta decay antineutrinos are too low to create a significant contribution to the background in water Cherenkov detectors.  

Accidental coincidence backgrounds arise when exactly two physically independent interactions appear close in time 'accidentally', mimicking the time-correlated antineutrino signal. These are also referred to as uncorrelated backgrounds. The remaining backgrounds in the above list are correlated backgrounds, meaning the same physics process produces both the prompt and delayed events, with a time distribution similar to that of the antineutrino-induced  positron-neutron signal.

The accidentals rate does not depend on depth but does depend on the composition of the detector and environs, and on factors such as the shielding distance of the inner volume from the edge of the detector. The accidental background rate also depends on energy threshold and vertex position in the detector, the vertex resolution and the degree of fiducialization. Fiducialization refers to the definition of a good or 'fiducial' central region, away from the PMTs and detector walls, based on the reconstructed location of the event in the water volume.  This location is reconstructed by comparing the arrival times of individual Cherenkov photons at the hit PMTs. Even after fiducialization, ambient radioactivity from the detector edge (PMTs and walls) may still contribute to the accidentals rate via gamma straggling and vertex misreconstruction. The simulations described below provide estimates of the accidentals rates before and after fiducialization.  

Spallation caused by unvetoed or 'punch-through' muogenic fast neutrons  generates neutron pairs in the target, which subsequently thermalize and capture with a time interval and energy ranges characteristic of reactor antineutrinos. The main concern is muons interacting outside of the detector which do not produce a veto signal. We note that neutron-induced proton recoil events an important background in liquid scintillator detectors, are not a problem in water as nearly all recoiling protons are below the Cherenkov threshold. This does reduce the direct contribution of high energy neutron recoil backgrounds in the target region, but renders these fast neutron recoil events invisible in the veto region. 

\begin{table}
\caption{Expected rate of neutrons ($n$) at a depth of 1400 m.w.e. emerging from the side-walls. }
\begin{tabular}{ c  c c c c  } 
\hline\hline
$E_{n}$ above & XZ Walls & YZ Walls 	& All Side Walls & Flux\\
			& (Hz)	& (Hz)		&(Hz)		& $n\cdot$day$^{-1}$\\
\hline
 $>$1 (MeV)		& 5.77e-02 		& 4.36-02 		& 2.03e-01 &17,500\\
 $>$10 (MeV)		& 4.02e-02 		&3.04e-02 	&1.41e-01 &12,200\\
$>$100(MeV)		& 8.34e-03 		& 6.30e-03 	& 2.93e-02&2,530\\

\hline\hline
\label{table:FastNeutronFlux}
\end{tabular}
\end{table}

Long-lived muogenic radionuclides, such as $^{9}$Li, can in the course of decaying generate paired signals with time structures and energy depositions that are similar to the antineutrino signal. The long lifetimes make direct vetoing of these radionuclide backgrounds difficult, though a veto may be effected for a portion of the signal. Because this correlated background has not yet been directly measured in water, we rely on indirect estimates from related experiments in order to set bounds on the rates. This results in a large uncertainty in the radionuclide background estimate. Even with this uncertainty,  we show here that the baseline WATCHMAN design can easily attain the required sensitivity goals for the reactor signal.

The WATCHMAN  collaboration is now in the process of making direct measurements of the rate of correlated event production from muogenic radionuclides, using a few ton dedicated underground Gd-H2O detector known as WATCHBOY. The collaboration has also fielded a high energy neutron spectrometer, the Multiplicity and Recoil Spectrometer (MARS) at multiple depths in an underground mine, in order to measure the fast neutron background. Upon completion of the measurement campaigns, the WATCHBOY and MARS data will together be used to reduce the uncertainty in the background estimates. 

Absent direct measurements, we used two approaches to estimate the rate of $^{9}$Li for WATCHMAN. The first approach was to scale the estimated radionuclide rate from the KamLAND experiment \cite{Abe:PRC81}, a scintillator-based experiment in which the primary radionuclide production target is carbon.  The second was to scale an upper limit on radionuclide production in  the Super-Kamiokande detector \cite{Li:PRD73}, a water-based  experiment like WATCHMAN, in which the primary radionuclide production target is oxygen. Differences in detection efficiencies, overburden, and fiducial mass were accounted for in the scaling process. For the KamLAND scaling, the production rate of radionuclides on carbon was assumed to be identical to that for oxygen. 
  
Real antineutrinos from other reactors also form a small part of the background for WATCHMAN. These are calculated using the IBD efficiency as derived from the simulation, based on the locations of all known nearby reactors. This background is not significant for our measurements at the Morton site.
 
We modeled the accidental coincidence rate, muogenic fast neutron and radionuclide rates in WATCHMAN, assuming our baseline design. If no backgrounds were present, our simulations indicate that WATCHMAN would achieve a remarkably high intrinsic signal detection efficiency of $65 \%$.  This is indicative of the ultimate performance of the detector as it scales to larger masses, since the accidentals backgrounds will fall off quickly as the detector mass grows. This rapid falloff with increasing target volume is due to the fact that the thickness of water necessary to suppress wall and PMT backgrounds grows much more slowly than the target volume as the detector mass is increased. However, as shown here, the efficiency of  the demonstration WATCHMAN detector is necessarily lower than $65 \%$, since residual backgrounds must be suppressed in this relatively small detector, at some (still relatively modest) cost in signal efficiency. 

The models used for each class of background in the simulation were as follows: 
\begin{enumerate}

\item The rate of di-neutrons from cosmogenic fast-neutrons was evaluate from Mei and Hime predictions \cite{Watchman:MeiHime}.  The neutron flux estimated for the four walls of a cavern the size of the Davis cavern are shown in Table \ref{table:FastNeutronFlux}. With our baseline one meter thick veto and fiducial skin, the vast majority of the fast neutrons that reach the detector are either caught by the detector veto or reconstruct outside of the fiducial volume. Our study shows that out of a incident rate of 17,500 neutrons/day, only $1.3\pm0.3$ neutron/day will create a prompt and delayed pair within the fiducial volume.  

\item the U/Th/K activity from the PMT glass (341 ppb/1.33 ppm/260 ppm); gammas from the U/Th chains were isotropically generated at the PMT were found to reconstruct within the Fiducial volume less than 4\% of the time.

\item the background arising from PMT dark noise was taken into account in two steps. First, PMT dark noise was added in the time window of the event to properly reproduce the triggering conditions. Second, independent toy Monte Carlo studies were made of triggers arising from statistical fluctuation as a function of the trigger time window.  The rate of PMT dark noise induced backgrounds was found to be negligible. 

\item the detector response to $^{9}$Li was evaluated by generating a positron from a beta-decay spectrum and an associated decay neutron at a vertex distributed throughout the detector volume and with the correct decay lifetime. The detector is more sensitive to $^{9}$Li than IBD in part due to the higher prompt energy of the positron. The detection efficiency could be as high as 85\% depending on the triggering conditions. However, two methods are available to reduce the rate of this background.  FIrst  the time veto after a high-energy muons may be increased (particularly for showering muons, which are more likely to produce radio-nuclides). Second, events can be removed that are near a reconstructed muon track, while allowing valid events elsewhere in the fiducial region. The results shown in this section do not have these rejection techniques applied and are therefore conservative.   

\end{enumerate}

The estimates for these backgrounds as a function of detector threshold are shown in Figure 5. 
\begin{figure}
\includegraphics[width=0.48\textwidth]{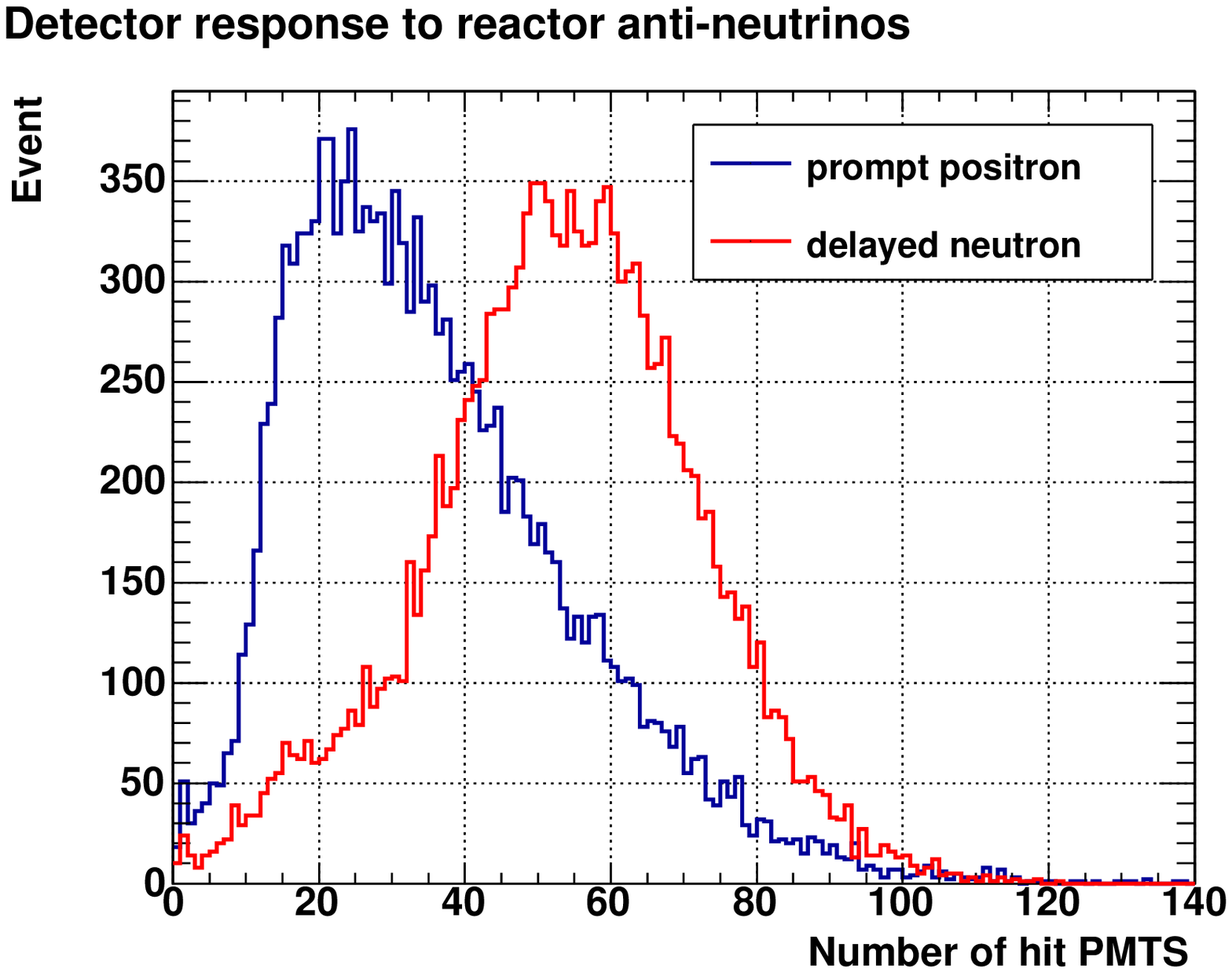}  
\includegraphics[width=0.48\textwidth]{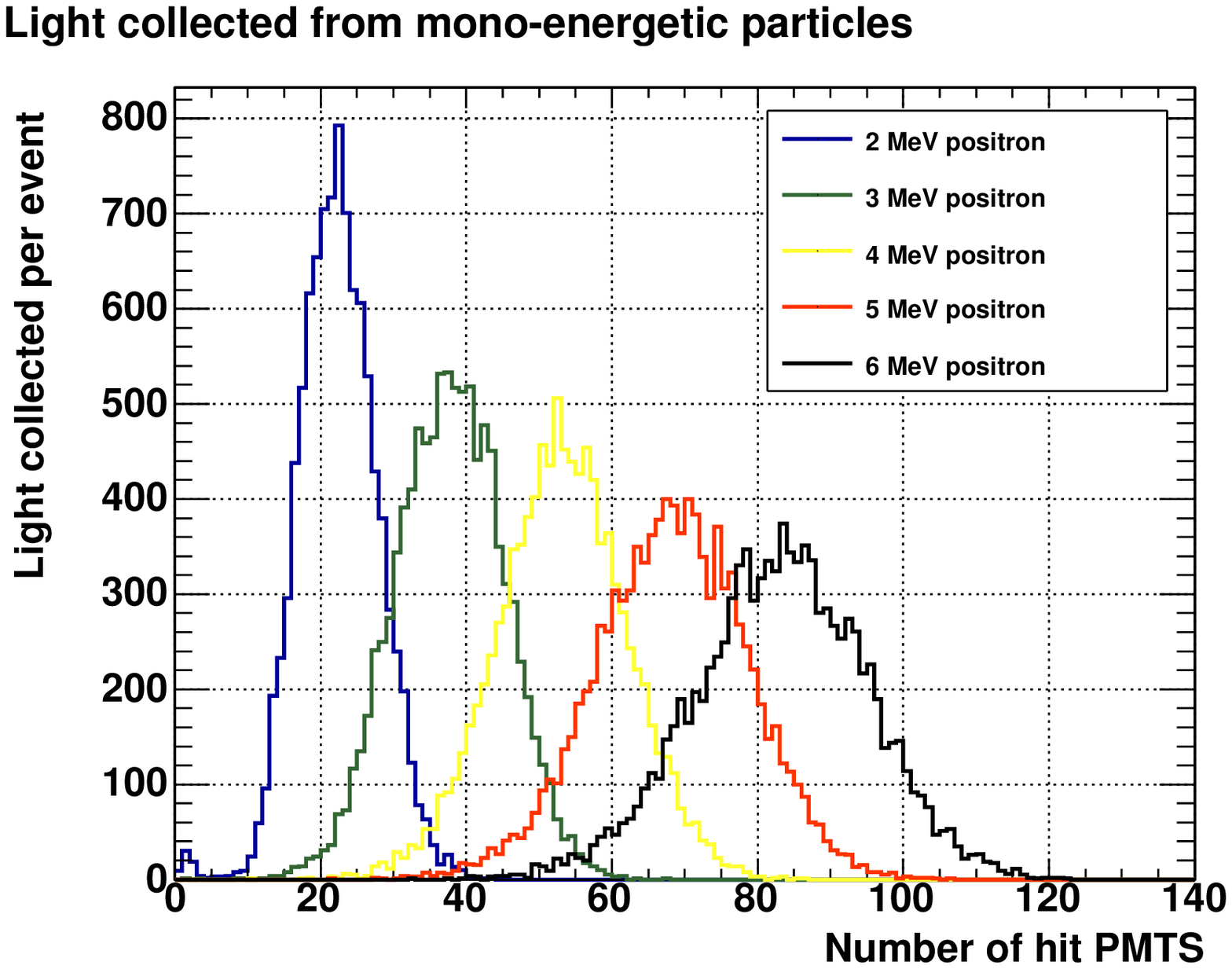}
\caption[]{Detector response to reactor anti-neutrino inverse beta decay on the left.  On the right is detector response to mono energetic positrons (in kinetic energy).}
\label{SBSIM:Signal}
\end{figure}
\begin{table}[h]
\label{SimTable}
\caption{Signal and background estimates for the Standard glass PMT option and the PMT installed 1 meter away from the fiducial volume.  The best and worst case scenario are evaluated where the signal to background ratio is maximized as is shown in Figure~5, right.} 
\begin{tabular}{ l  r r} 
\hline
&Best Case (low $^{9}$Li expectation)& Worst Case  (high $^{9}$Li expectation)\\
\hline
~~\bf{IBD Rate} 		\bf{(day$^{-1}$)}		&   	\bf{4.3}  	&  	\bf{4.6}   \\
~~IBD Efficiency	(\%)			&   22  			& 24  \\
\hline
{\bf Background Rate} 	\bf{(day$^{-1}$)}	&   \bf{1.8}   	& \bf{6.5}   \\
~~Dark noise ($\leq$ 4.5 kHz) (day$^{-1}$)	&$\ll 0.1$ & $\ll 0.1$  \\
~~Accidentals coincidence	(day$^{-1}$)	&    0.2    	& 0.3     \\
~~Fast-neutron (day$^{-1}$) 				&   1.3     	& 1.3      \\
~~Radio-nuclide ($^{9}$Li) (day$^{-1}$)	&   0.3    	& 4.9     \\
\hline 
\label{SimTable}
\end{tabular}
\end{table}

\begin{figure}
\includegraphics[width=0.96\textwidth]{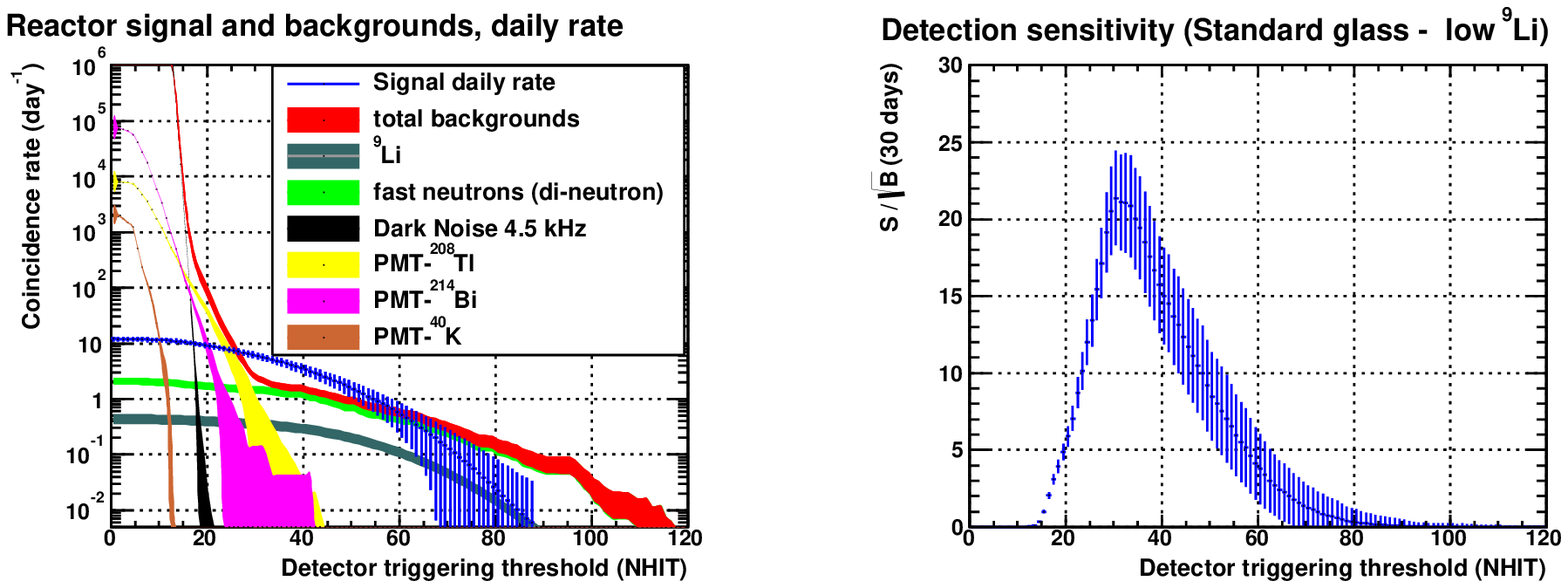}  
\includegraphics[width=0.96\textwidth]{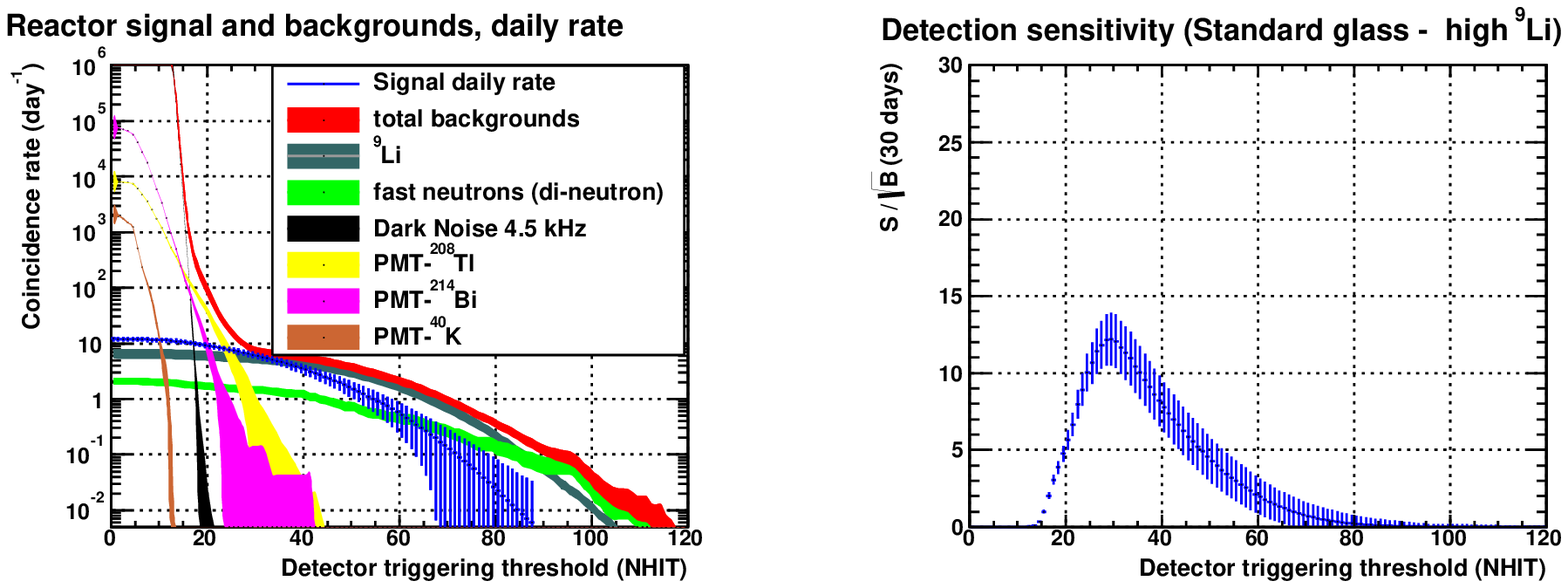}  
\caption[]{Signal and background as a function of the detector threshold and the detection sensitivity ($S/\sqrt{B}$) for both the low and high estimates of radio-nuclide production. No muon tagging was assumed to reduce this radio-nuclide rate, however the low-$^{9}$Li represent a scenario where the radio-nuclide background rates are small compared to the di-neutron rates from fast-neutron and therefore reflect the optimal improvements that could be achieved from muon-tagging of radio-nuclide.}
\label{SBSIM:Fig}
\end{figure}

The analysis criteria were that the events reconstruct within the 1-kton fiducial volume and that the time difference between the prompt-positron and delayed-neutron be less than 100 $\mu$s. This short capture time between prompt and delayed events allows both a lowering of the energy threshold and a significant reduction in backgrounds. The accidentals coincidence rate ($C$) per day can be estimated from the singles rate ($S$) per second with the power law $C= 8.6\cdot S^{2}$. The probability, evaluated by simulation, that two events are at most two meters from each other in the fiducial volume is ($2.3\pm$0.5)\%. This reduction results in a coincidence rate in the fiducial volume of $C=(0.20\pm0.05)\cdot S^{2}$. 

In summary, we have shown that even with conservative assumptions about backgrounds,  the baseline WATCHMAN design can easily accomplish its main nonproliferation goal of high-confidence detection of the reactor IBD signal using two 30 day periods of reactor on ( $100\%$ power) and reactor off data.The simulation package developed for the reactor analysis has been extended in a consistent way to permit analyses of other physics interactions, which are described in later sections of this white paper.

\section{Physics and Detector Research and Development Goals}

The WATCHMAN physics program begins with goals that are achievable with the baseline Gd-H2O Water Cherenkov detector. The main physics goals in the first phase of the program are:

\begin{enumerate}

\item world-class supernova sensitivity, he first such capability in the U.S. in two decades, including the ability to separately identify electron neutrino and antineutrino signals, to recover directional information, and to potentially provide an early warning to experiments such as LBNE and LIGO that seek to use a neutrino pre-trigger to arm their detectors for supernova physics;

\item competitive sensitivity to sterile neutrinos, using a compact low energy neutrino beam provided by the ISODAR program

\item unique sensitivity to non-standard neutrino interactions, using the ISODAR neutrino beam.

\end{enumerate}

In a second phase, using water-based scintillator or oil-based scintillator, the main goals are:

\begin{enumerate}

 \item Enchanced sensitivity to sterile neutrinos and non-standard neutrino interactions,  using the ISODAR  beam; and 

\item increased efficiency and spectral resolution for supernova antineutrinos.  

\end{enumerate} 

WATCHMAN's sensitivity to the neutrino mass hierarchy was examined and found to be limited at this standoff, regardless of the target fill. 

In addition to the physics goals, WATCHMAN is an excellent U.S.-based test-bed for technological advances that are relevant for future large scale water detectors. The main technological advance, demonstration of low energy antineutrino sensitivity using Gd-doped water, is already the main goal of the WATCHMAN nonproliferation program.  In addition, new technologies that can be fielded in WATCHMAN include, Large Area Picosecond Photodetectors, high quantum efficiency PMTs, U.S.-manufactured PMTs, water-based liquid scintillator, Winston cones and wavelength shifting plates can all be tested in WATCHMAN. This will help ensure U.S. involvement and competitiveness as large-scale water-based antineutrino detectors are planned and built worldwide. 

In the following sections we discuss each of the physics goals for the water detector and the follow-on detectors.

\newcommand{\nuebar}{\mbox{$\bar{\nu}_e$} }
\newcommand{\gad}{\mbox{Gd$_2$(SO$_4$)$_3$} }

\subsection{Supernova}

Supernova neutrinos carry unique information about one of the most dramatic 
processes in the stellar life-cycle, a process responsible for the production 
and dispersal of all the heavy elements (i.e., just about everything above helium) in 
the universe, and therefore a process absolutely essential not only to the look and feel of the 
universe as we know it, but also to life itself. As a gauge of the community's level of interest in these 
particular particles, it is worth noting that, based upon the world sample 
of twenty or so neutrinos detected from SN1987A, 
there has on average been a theoretical paper published once every 
ten days -- for the last three decades!  This makes it all the more
surprising that the US has not had a supernova neutrino detector of its own in 
operation since the end of the IMB experiment (which recorded eight neutrinos from 
SN1987A) in 1991. WATCHMAN will be the first US detector since IMB to have world-class sensitivity to supernova neutrino and antineutrinos. 

The next time a Milky Way core collapse supernova goes off, an event 
expected to occur every 30 years or so~ \cite{0004-637X-778-2-164}, it would be extremely desirable to have 
a sizable gadolinium-doped water Cherenkov detector like WATCHMAN in operation 
when the resulting neutrino wave sweeps across the planet.  This is primarily because the most 
copious supernova neutrino signal by far ($\sim$88\%) comes from inverse beta events. 
They are only produced by one of the six 
species of neutrinos and antineutrinos which are generated  
by a stellar collapse, and so if we could tag them 
individually by their follow-on neutron captures then we could  
extract the \nuebar time structure of the burst precisely, gaining
valuable insight into the inner dynamics of the explosion.  
What's more, we could then subtract them away from the more subtle 
non-\nuebar signals, uncovering additional information that would
otherwise be lost during this once-in-a-lifetime happening. 

Due to its unique capability to immediately identify a galactic supernova
as genuine {\em while the neutrino wave is still passing through the
Earth,} WATCHMAN will be able to instantly alert both the astronomical
community and other projects that a core collapse is currently underway.
Major physics experiments such as LBNE, LIGO, and IceCube -- whose
challenging SN analyses will likely require matching extremely complex,
exceedingly noise-like, and/or exceptionally sub-threshold signals to a
precise external start time -- would  benefit greatly from a realtime
alert from WATCHMAN.  The knowledge gained in the precious few hours
between the arrival of the neutrinos (which are generated first) and the
arrival of the supernova's first light (the so-called shock breakout) will
serve to guide and optimize the subsequent multi-wavelength observations
of the dying star.  In particular, the physical proximity of WATCHMAN to LBNE means that rapid (few ms) signaling of a supernova to the Long Baseline Neutrino Experiment's \cite{Akiri:2011dv} trigger system  is possible. This may prove to be useful in simplifying or strengthening LBNE's supernova watch capability, with potential attendant cost savings and improved physics reach. Additional work with the LBNE collaboration - of which
several of the present authors are members -  will seek to quantify these improvements or savings.

Table~\ref{tab:snsig} shows the expected signals in WATCHMAN for a galactic type II supernova. 
For a stellar collapse 10,000 light years distant from earth (somewhat less than halfway to the galactic center) 
there will be a total of about 4,000 events seen in WATCHMAN.  As one can see, the inverse beta events 
dominate the expected signals.  However, there is valuable information to be
gained from the other signals, information which would be largely inaccessible without
neutron tagging from dissolved \gad.

\begin{table}
\begin{center}
\begin{tabular}{|c|c|c|}
\hline
Neutrino & Percentage of & Type of\\  
Reaction & Total Events & Interaction \\ \hline 
$\overline{\nu}_e + p        \rightarrow  n + e^+$ & 88 & Inverse Beta \\
$\nu_e +  e^-                \rightarrow  \nu_e + e^-$ & 1.5 & Elastic Scattering \\
$\overline{\nu}_e +  e^-     \rightarrow  \overline{\nu}_e + e^-$ & $<$1 & Elastic Scattering \\
$\nu_x +e^-                  \rightarrow  \nu_x + e^-$ & 1 & Elastic Scattering\\
$\nu_e +  ^{16}O             \rightarrow  e^{-} + ^{16}F$  & 2.5 & Charged Current \\
$\overline{\nu}_e +  ^{16}O  \rightarrow  e^{+} + ^{16}N$ & 1.5 & Charged Current \\
$\nu_x +^{16}O               \rightarrow \nu_x +O^{*}/N^{*}+\gamma$ & 5 & Neutral Current \\ 
\hline
\end{tabular}
\caption{Breakdown of supernova neutrino events expected in WATCHMAN from
a galactic supernova.  Oscillations are taken into account.
$\nu_x$ indicates the total interactions of $\nu_{\mu}$, $\nu_{\tau}$, and
their antineutrinos.}
\label{tab:snsig}
\end{center}
\end{table}

For example, being able to tag the \nuebar events would 
immediately double WATCHMAN's pointing accuracy back to the progenitor
star.  This is merely the result of statistics, since our elastic scatter 
events (about 3\% of the total) would no longer be sitting on a large 
background in angular phase space~\cite{PhysRevD.68.093013}.  WATCHMAN would be the only 
detector in the world other than Super-Kamiokande -- and the only detector in the Western Hemisphere -- 
with neutrino pointing capability. Reducing the error on this quantity by a factor of two, roughly
from 10 degrees to 5 degrees,  
via neutron tagging would reduce the amount of sky to be searched by a factor of four.  
This could prove quite important for the wide-field
astronomical instruments which would be frantically attempting to see the 
first light from the supernova.

At the same time, this event-by-event subtraction could allow identification 
of the initial electron neutrino pulse from the neutronization of the infalling 
stellar matter, a key and as yet unobserved input in understanding supernova dynamics.

What's more, the neutral current events, which may be easily identified by 
their mono-energetic gammas between 5 and 10 MeV once the \nuebar events 
are subtracted, are very sensitive to the temperature of the burst and 
the subsequent neutrino mixing~\cite{PhysRevLett.76.2629}. 

With a neutrino energy production threshold of 15.4 MeV, the weakly
backward-peaked charged current 
events are even more sensitive to the burst temperature and the subsequent 
mixing~\cite{PhysRevD.36.2283} than the NC events.  As there are no other good $\nu_e$ detectors 
currently in operation, WATCHMAN's charged current events, if distinguishable 
from the \nuebar background, would provide one of the only sources of data on 
supernova $\nu_e$'s.

If the exploding star was big and rather close (say, like Betelgeuse at 600 light years) 
we would get an early warning of its impending collapse~\cite{Odrzywolek2004303}.  Approximately
a week before exploding, the turn-on of silicon fusion in the core would
raise the temperature of the star sufficiently that electron-positron annihilations
within its volume would begin to produce \nuebar just above inverse beta threshold.  
The sub-Cherenkov positrons would be invisible, but in WATCHMAN the captures of the 
resulting neutrons on \gad would result in a sudden and continuing 
increase of our usual low energy singles rate. 

As early as one day before collapse we would 
see a several sigma excursion in our low energy singles rate.  The continuing increase 
in singles rate would clearly indicate a coming explosion, ensuring that we did 
not intentionally turn off WATCHMAN for calibration or maintenance and thereby 
miss the coming explosion. 

In addition, a gadolinium-enriched WATCHMAN would be sensitive to quite late 
black hole formation following a supernova explosion within our galaxy, 
since the distinctive coincident inverse beta signals from the cooling phase 
could be distinguished from the usual singles backgrounds. An abrupt cutoff 
of these coincident signals occurring even several minutes after the main burst 
would be the conclusive signature of a singularity being born. Direct observation
of such an event would clearly be of great value, especially when correlated with
electromagnetic signals from X-ray or gamma-ray observatories as well as gravitational 
wave observations.

Finally, WATCHMAN would allow us to be absolutely 
certain that a supernova was occurring the moment the data started to arrive.
This is because the distinctive correlated \nuebar event pairs, with the
positron and the neutron capture separated in time by only tens of microseconds, 
would in general be well separated from other supernova events, 
which at for an explosion at 10,000 light years would come in every millisecond or so at the 
height of the burst.  This would yield a clear ``heartbeat''-like time structure 
in the time plot of the events -- something that could be easily spotted online.
Of course, if the supernova was close enough that these event pairs were overlapped 
by subsequent SN signals then we would also know for sure that something 
exciting was happening, and in such a near-field case we may very well have been 
waiting for the burst anyway, due to the early warning from silicon burning.  
Either way, the surer we are that a supernova has just occurred, the faster 
we should be able to get word out to the community.  

So, from extracting the neutronization signal, to deconvolving the main burst, to 
pointing back at the progenitor star, to 
observing late black hole formation, to eventually announcing the event to the world, 
having \gad in WATCHMAN would positively impact just about every 
physics topic connected to the detection of a galactic supernova.
Much of this physics would be buried in background, degraded in precision, or 
wholly inaccessible otherwise.  In summary, the key benefits are:

\begin{enumerate}

\item Neutron tagging of the inverse beta events would allow the de-convolution of a galactic 
supernova's various signals, which in turn would allow much more detailed interpretation of 
the physics of the burst.  This will result in significantly improved pointing back to the exploding star.

\item Early warning (hours before the arrival of the supernova neutrino 
wave) of large, relatively nearby supernovas would be possible via the observation of 
silicon-powered fusion in the dying stellar core.

\item Gadolinium would allow very late time black hole formation -- minutes after the initial 
explosion -- to be observed.  This critically important signal could otherwise be hidden by background events.

\item The ``gadolinium heartbeat'' of prompt events rapidly followed by delayed events in about the 
same locations would make the arrival of a genuine neutrino burst instantly identifiable, vital for getting 
word out to the rest of the world in a timely fashion.

\end{enumerate}

%
%
%
%
%
%
%
%


\subsection{Sterile Neutrino Search}

\label{sterile}

In recent years improved data and models of reactor antineutrino emission led to an increase in the predicted antineutrino flux from fission reactors, which in turn led to an apparent deficit in observed flux over many separate experiments, relative to the new prediction~\cite{PhysRevC.83.054615},~\cite{PhysRevC.84.024617}.  This deficit has been referred to as the reactor anomaly.  Short baseline accelerator experiments such as LSND~\cite{PhysRevLett.77.3082},~\cite{PhysRevD.64.112007} and MiniBooNE~\cite{PhysRevLett.110.161801} have also reported anomalous results that are difficult to reconcile with the standard three-flavor neutrino model. This is sometimes referred to as the accelerator anomaly.  A popular (though by no means unique \cite{PhysRevC.84.024617}) interpretation of these results is that another oscillation channel exists, with at least one sterile (undetectable) neutrino and relatively large mass eigenstate of at least $\sim 1 eV$.  

The proposed Isotope Decay At Rest (IsoDAR) experiment ~\cite{PhysRevLett.109.141802} will search for evidence of short baseline sterile neutrino oscillation using an intense flux of  $<E> =6.4$~MeV electron antineutrinos resulting from beta decay-at-rest of $^{8}$Li.  The electron antineutrino source will need to be located next to a  kiloton-scale antineutrino detector capable of sufficient energy and position resolution to reveal oscillations in $L/E$, where $L$ is the source-detector distance and $E$ is the antineutrino energy.  

One of the first sites considered for the IsoDaR beam experiment was the KamLAND detector. KamLAND's position and energy resolution are more than adequate to observe oscillations if the missing mass eigenstate is $\sim 1 eV$.  Here we examine the sensitivity of the WATCHMAN detector  to sterile neutrinos using the IsoDAR beam. Of course, there are important practical considerations related to the feasibility of deploying the beam in the experimental hall. Very preliminary studies of the feasibility of deployment based on the cavern layout have been encouraging, though significant engineering work would be necessary to accommodate the IsoDAR beam. In this white paper, we set aside these considerations,  focusing on the sensitivity of WATCHMAN detector to the physics signal, assuming the beam is successfully deployed. 

\begin{figure}[h!]
\includegraphics[width=1.0\textwidth]{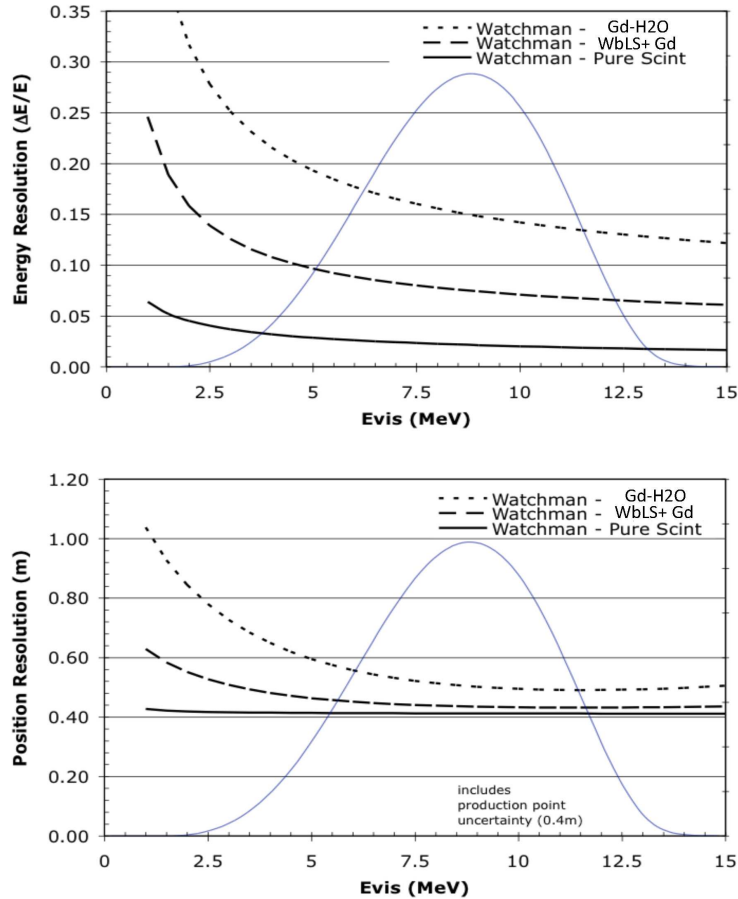}
\caption{The antineutrino flux spectrum of interacting antineutrinos in WATCHMAN.  An estimated $6.26 \times 10^3 $~antineutrino interactions can be generated over a five-year period with IsoDAR at the proposed beam position (16 meters from the center of the detector). The energy and position resolutions for Gd-doped water, Gd-doped water based liquid scintillator and pure scintillator are also given as a function of energy. }
\label{EandPosResolution}
\end{figure}

\begin{figure}[h!]
\includegraphics[width=1.0\textwidth]{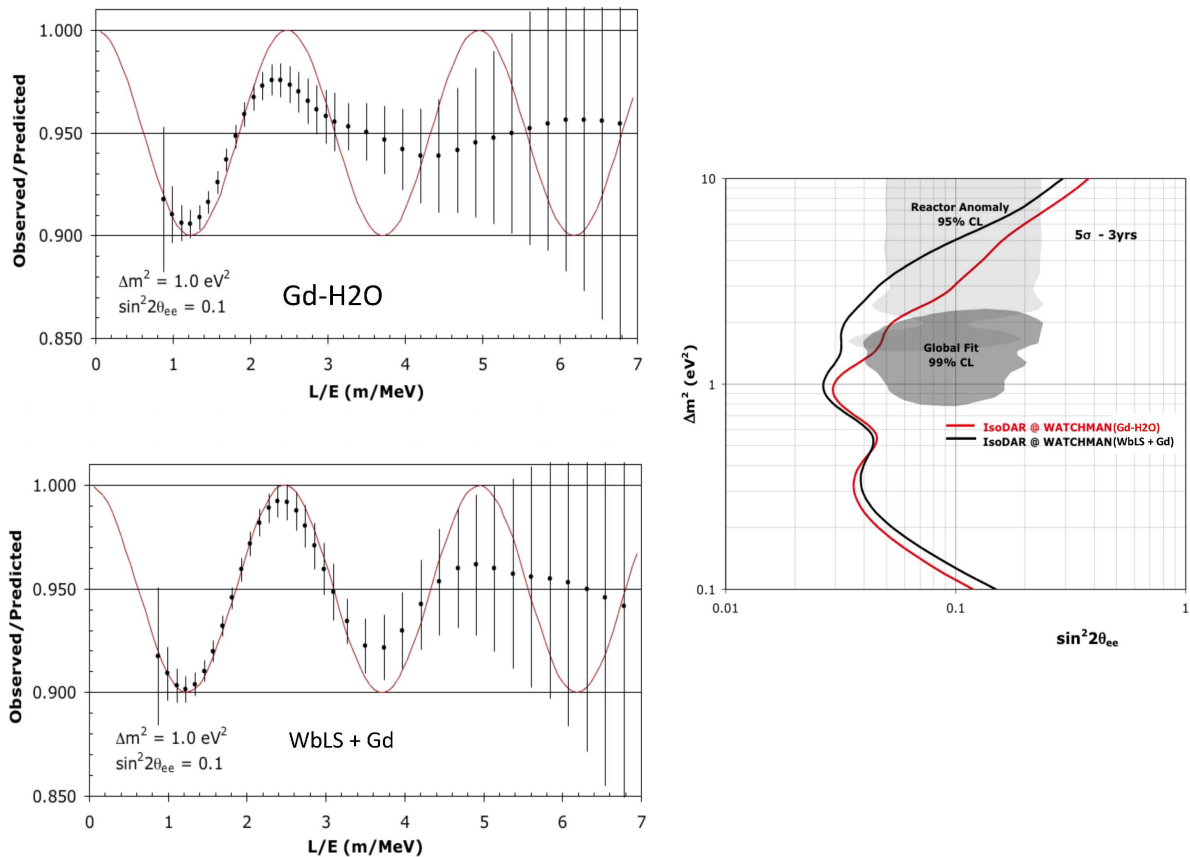}
\caption{Left: The sensitivity of the WATCHMAN detector to $L/E$ (baseline distance $L$ divided by antineutrino energy $E$), for two different detection media - water (upper plot) and light scintillator (lower plot). Right: The sterile neutrino oscillation parameter space  excluded by WATCHMAN after 3 years of run time, assuming Gd-doped water  (red) and Gd-doped water-based liquid scintillator (black). } 
\label{fig:WatchmanSensitivity}
\end{figure}

The key difference between the KamLAND and WATCHMAN detectors is the detection medium, which was liquid scintillator in KamLAND, and will be gadolinium-doped water in WATCHMAN (in its first phase).  In this case, both energy and position resolution will be lower than for KamLAND, due to the significantly lower photon production from Cherenkov emission compared to scintillation. However, we anticipate that the resolution in WATCHMAN will be good enough to exclude the region of oscillation parameter space most favored by the reactor and accelerator experiments to date (see Figure~\ref{fig:WatchmanSensitivity}) or indeed discover an oscillation if it is there. 

The IsoDAR beam is expected to produce approximately $1.3 \times 10^{23}$ antineutrinos over a five-year period at an average energy of $6.4$~MeV. The energy spectrum of the resulting antineutrino interactions is shown in Figure~\ref{EandPosResolution}. It is slightly higher in energy due to the cross section of the inverse beta decay reaction on protons, $ \bar{\nu} + p \rightarrow e^{+} +n$, which increases with the  square of the antineutrino energy.  Just as for reactor antineutrino experiments, the positron energy ($E_{e^+}$), is related to the antineutrino energy ($E_{\nuebar}$), by

\begin{equation}
E_{e^+}=E_{\nuebar} -1.8 MeV
\end{equation}

The resulting positron energy distribution peaks at approximately $7 MeV$.  As we show here,  despite the absence of scintillator in WATCHMAN Phase I, the $\sim 7 MeV$ positrons are detectable with sufficient resolution in the Gd-H2O target to perform a sensitive search for oscillations. 

An IsoDAR deployment at a distance of $L = 16m$ from WATCHMAN will have maximal sensitivity to mass squared difference near $\Delta m^{2} \sim 1 eV^{2}$, the region most favored by the reactor and accelerator anomalies.  Figure~\ref{fig:WatchmanSensitivity} shows the expected sensitivity of IsoDAR deployment next to WATCHMAN assuming a 3 year run.  The sensitivity for WbLS is also shown, where the medium is assumed to have a light yield of a few hundred scintillator photons generated per MeV. 

The water-based WATCHMAN detector was assumed to have photocathode coverage of $40\%$, using 12-inch high quantum efficiency PMTs.  Approximately 10 photoelectrons per MeV can be expected from positrons in water.  The sensitivity plots in Figure~\ref{fig:WatchmanSensitivity} were based on simulations done for large water based detector proposals such as LBNE, which in turn grew out of well tested and tuned simulations from the Super-K detector.

In generating the sensitivity plots, we used the backgrounds in WATCHMAN that were calculated  for the reactor monitoring task, as summarized in section~\ref{sec:sbsims}.  Based on those simulations, we expect that WATCHMAN will detect approximately 5 IBD events per day from the nearby Perry reactor.  Backgrounds from both cosmogenic radionuclide production and fast neutrons from the rock are expected at about 6 per day.  Backgrounds from the rock or from the PMT structures inside the detector tend to be subdominant due to the requirement that IBD detection depends on the coincidence detection of both the positron and delayed neutron. Neutron and gamma-ray backgrounds from the IsoDAR beam itself will be shielded by rock between the beam and the detector.  We assign a total background rate of $\sim 20$ events per day from  the reactor and all other backgrounds.  

We conclude that a sensitive sterile neutrino search can be conducted with the WATCHMAN detector in its first and follow-on phases, using both the Gd-H2O and Gd-WbLS targets.

\subsection{Non-Standard Neutrino Interactions}

\label{sec:NSI}

As described by \cite{Conrad:2013sqa}, a sensitive search for non-standard neutrino interactions can be performed by collecting a sufficiently large number of antineutrino electron scattering events 
($ \bar {\nu} + e^{-} \rightarrow   \bar {\nu} + e^{-} $). In both the pure Gd-H2O and scintillator mode, the WATCHMAN detector provides a large target region that can be used for this purpose, with the antineutrinos coming from the ISODAR beam. Examples of non-standard interactions include effects on the cross-section arising from heavy partners of the known light neutrinos, or new gauge bosons that couple only to neutrinos. In this section we discuss the sensitivity of the WATCHMAN Gd-H2O and scintillator options to non-standard neutrino interactions.  The IsoDAR source characteristics, and the beam and detector geometry assumed for the calculations here are the same as described in section~\ref{sterile}.  However, in this section, we examine Gd-H2O and oil-based scintillator options, rather than water-based scintillator. 

Following \cite{Conrad:2013sqa}, the modified Standard Model cross section incorporating potential non-standard couplings is written as: 

 \begin{equation}  
 \label{nsi}
\frac{d \sigma (E_{\nu},T)}{dT} = \frac{2 G^{2}_{F}m_{e}}{\pi} \left(\tilde{g}_R^2+\tilde{g}_L^2 (1-\frac{T}{E_\nu})^2+\tilde{g}_R \tilde{g}_L\right)
 \end{equation}

Here $\tilde{g}_{R} = g_{R}+ \epsilon_{ee}^{eR} $and  $\tilde{g}_{L}  = g_{L}+ \epsilon_{ee}^{eL} $ are the  Standard Model couplings $g_{L(R)}$ , modified by the NSI parameters $ \epsilon_{ee}^{eR(L)} $ for right-handed and left-handed couplings. Flavor-violating couplings to muon and tau neutrinos are strongly restricted \cite{1126-6708-2003-03-011} and are not included in Eqn.~\eqref{nsi}.

We consider two cases: the WATCHMAN Phase 1 pure Gd-H2O target and the WATCHMAN Phase 3 scintillator target. We begin with the scintillator case, expected to be similar to KamLAND/IsoDAR. This case has already been studied extensively \cite{Conrad:2013sqa}.  We then consider the different treatment of backgrounds that would be required in a Gd-H2O detector.

\subsubsection{Sensitivity with a scintillator target}

Backgrounds divide into two categories, beam-related IBD events with a missed final-state neutron, and non-beam backgrounds arising from solar neutrino interactions, muon spallation, and ambient radiation from the detector and nearby rock.  In scintillator, the non-beam backgrounds can be measured $in~situ$ accurately by simply turning off the beam. We assume here one year of total beam off time will be devoted to this measurement.  The spectral shape of the IBD backgrounds can be measured by employing the neutron capture as a tag. The normalization of this background will depend on an accurate knowledge of the neutron tag inefficiency - obtained from calibrations. Gamma ray backgrounds from the detector wall components and from radon in the scintillator are scaled from the KamLAND/IsoDar prediction by relative fiducial volume. Cosmogenic radionuclides are expected to contribute significantly more to the background at WATCHMAN, since they are depth dependent (see Table~\ref{tab:NSIevents}), and WATCHMAN will be shallower (1430 m.w.e. versus 2700 m.w.e.). We assume that the bulk of radionuclide backgrounds are caused by showering (high energy) muons passing through the detector. Using a GEANT4 simulation of muon flux as a function of depth and energy, and using a 
representative sea level muon energy spectrum, the predicted rate of showering muons at WATCHMAN, relative to KamLAND is 2.6 times higher. At this rate, the 5 second KamLAND veto must be reduced to 2 seconds in order to maintain an equivalent live time.  We extrapolated the $^{11}$Be, $^{8}$Li and $^{8}$B rates from KamLAND/IsoDAR using the known isotope lifetimes (See Table~\ref{tab:LiBeacomRadNucs}) to find the rates at WATCHMAN. The WATCHMAN results shown in Table~\ref{tab:NSIevents} predict approximately 7030 electron scatter events within the WATCHMAN fiducial volume from an IsoDAR beam situated 16 meters from the center of the detector.

\subsubsection{Sensitivity with a pure Gd-H2O target}

In water, antineutrino-electron scattering events generate a Cherenkov cone that arises from the recoiling electron. The low light output compared to scintillator changes both the signal and background detection efficiencies and alters the strategy for isolating the signal. The presence of oxygen in the target, rather than carbon, increases the variety of radionuclide isotopes that may be produced via muon spallation. The exact isotopes and their yields, however, have never been measured in water. We rely on the values in Table~\ref{tab:LiBeacomRadNucs}), from \cite{Li:PRD73}, generated from a FLUKA simulation of the Super-Kamiokande detector. The expected event rates for each of the five most dangerous radionuclides were presented in Table~\ref{tab:NSIevents}.

\begin{table}[htbp]
\caption{The most dangerous long lived cosmogenic radionuclide backgrounds for the water-based Super-K target. The predictions were generated from a FLUKA study by \protect{\cite{Li:PRD73}}. The radionuclide isotopes of interest to the organic scintillator experiment include only a subset of these, $^{11}$Be, $^{8}$Li and $^{8}$B \protect\cite{Conrad:2013sqa}.}
\begin{tabular}{|c|c|c|c|c|}
\hline
    \textbf{Isotope}  &  \textbf{Half-life (s)} & \textbf{Decay Mode} & \textbf{Yield} $\bf{(\times 10^{-7} \mu^{-1} g^{-1} cm^{2} )}$   &    \textbf{Primary Process} \\
    \hline
    $^{16}N$ & $7.13$ & $\beta^{-} \gamma (66\%),  \beta^{-} (28\%)$ & $18$ & (n,p) \\
    $^{15}C$ & $2.449$ &  $\beta^{-} \gamma (63\%),  \beta^{-} (37\%)$ & $0.8$ & (n,2p) \\
    $^{11}Be$ & $13.8$ & $\beta^{-} (55\%),   \beta^{-} \gamma (31\%)$ & $0.8$ &  (n,$\alpha +$ 2p) \\
    $^{8}B$ & $0.77$ & $\beta^{+}$ & $5.8$ &         ($\pi^{+},\alpha + $ 2p $+$ 2n) \\
    $^{8}Li$ & $0.84$ &  $\beta^{-}$ & $13$ & ($\pi^{-},\alpha + ^{2}H$ $+$ p $+$ n) \\    
    
        \hline
 \end{tabular}
  \label{tab:LiBeacomRadNucs}
\end{table}    

\begin{table}[htbp]
\caption{The estimated elastic scattering signal and predicted backgrounds after 5 years running time for IsoDAR at WATCHMAN. Estimates for both the liquid scintillator and $Gd-H_{2}O$ WATCHMAN targets are given. Also given are background values assuming we use the cosine of the electron scattering angle with respect to the IsoDAR beam as a discrimination parameter.}
    \begin{tabular}{|c|c|c|c|}
\hline
    \textbf{Event Type}  &  \textbf{Liquid Scintillator} &  $\bf{Gd-H_{2}O}$ & $\bf{Gd-H_{2}O, cosine(phi) > 0.5}$ \\
    \hline
    \textbf{IsoDAR Elastic Scattering Signal} & \textbf{7030}  & \textbf{9010} & \textbf{9010} \\
    & & & \\
    Misidentified IBD & 1920  & 153500 & 38380 \\
    $^{208}Tl$ gamma rays & 1154  & 1154 & 289 \\
    Steel Support Gamma rays & 437  & 437 & 110 \\
    Rock Gamma rays & 1025 & 1025 & 256 \\
    $^{8}B$ Solar Neutrinos & 1700 &  2180 & 540 \\
    Radionuclide $^{8}B$ & 3155  &  4880 & 1220 \\
    Radionuclide $^{8}Li$ & 5595  &  12720 & 3180 \\
    Radionuclide $^{11}Be$ & 2840  &  3690 & 920 \\
    Radionuclide $^{15}C$ & & 2320 &  580 \\
    Radionuclide $^{16}N$ & & 75750 & 18940 \\
    \textbf{Total Backgrounds} & \textbf{17830} & \textbf{257660} & \textbf{64415} \\

\hline
 \end{tabular}
  \label{tab:NSIevents}
\end{table}

The signal efficiency depends strongly upon on the visible energy threshold, $E_{vis}$. This is necessarily higher in water than in scintillator, since the number of photons produced is small. In the present analysis, the gamma ray emissions from impurities in the water, the rock walls and detector walls were estimated by scaling directly from the IsoDAR NSI paper \cite{Conrad:2013sqa}, with the same energy threshold of $E_{vis} = 3$ MeV applied. The radionuclide, solar neutrino and misidentified IBD backgrounds (from IsoDAR and from the Perry reactor) have known and well defined energy spectra. Therefore no low energy threshold was applied for these backgrounds.

Beam-related backgrounds are expected to be higher in water than scintillator, since the efficiency for detecting the IBD final state neutron is significantly lower ($\sim 80\%$) than in scintillator ($99.75\%$, \cite{Conrad:2013sqa} with cuts optimized for IBD rejection). Since the cross section per atom and incident antineutrino is around ten times higher for IBD events compared with ES events, This background must be further suppressed to achieve a viable result in water.  For this purpose, we rely on the angular correlation between the reconstructed angular direction of the recoiling electron for ES events and the incoming antineutrino.

\begin{figure}
\begin{center}
\includegraphics[width=0.45\textwidth]{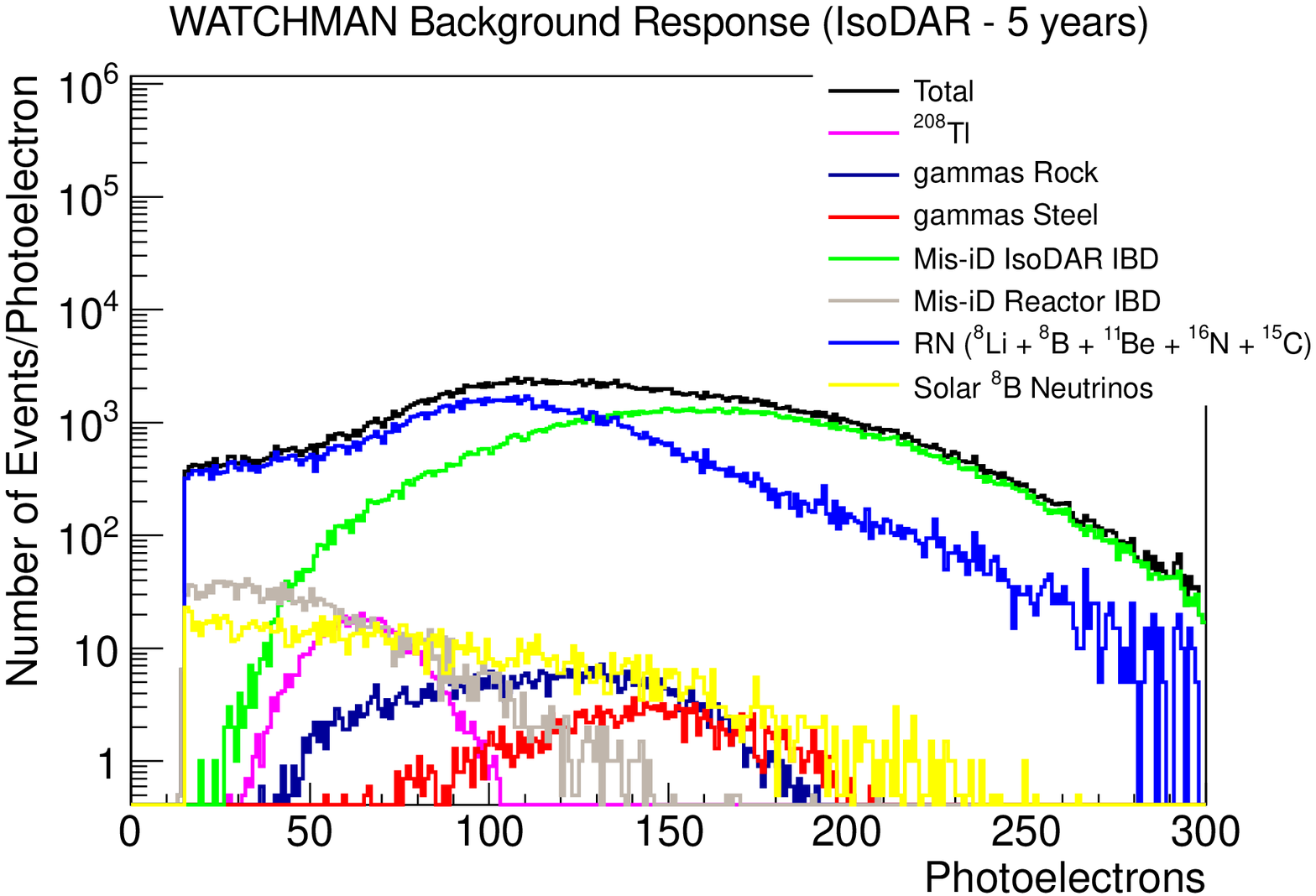}
\includegraphics[width=0.45\textwidth]{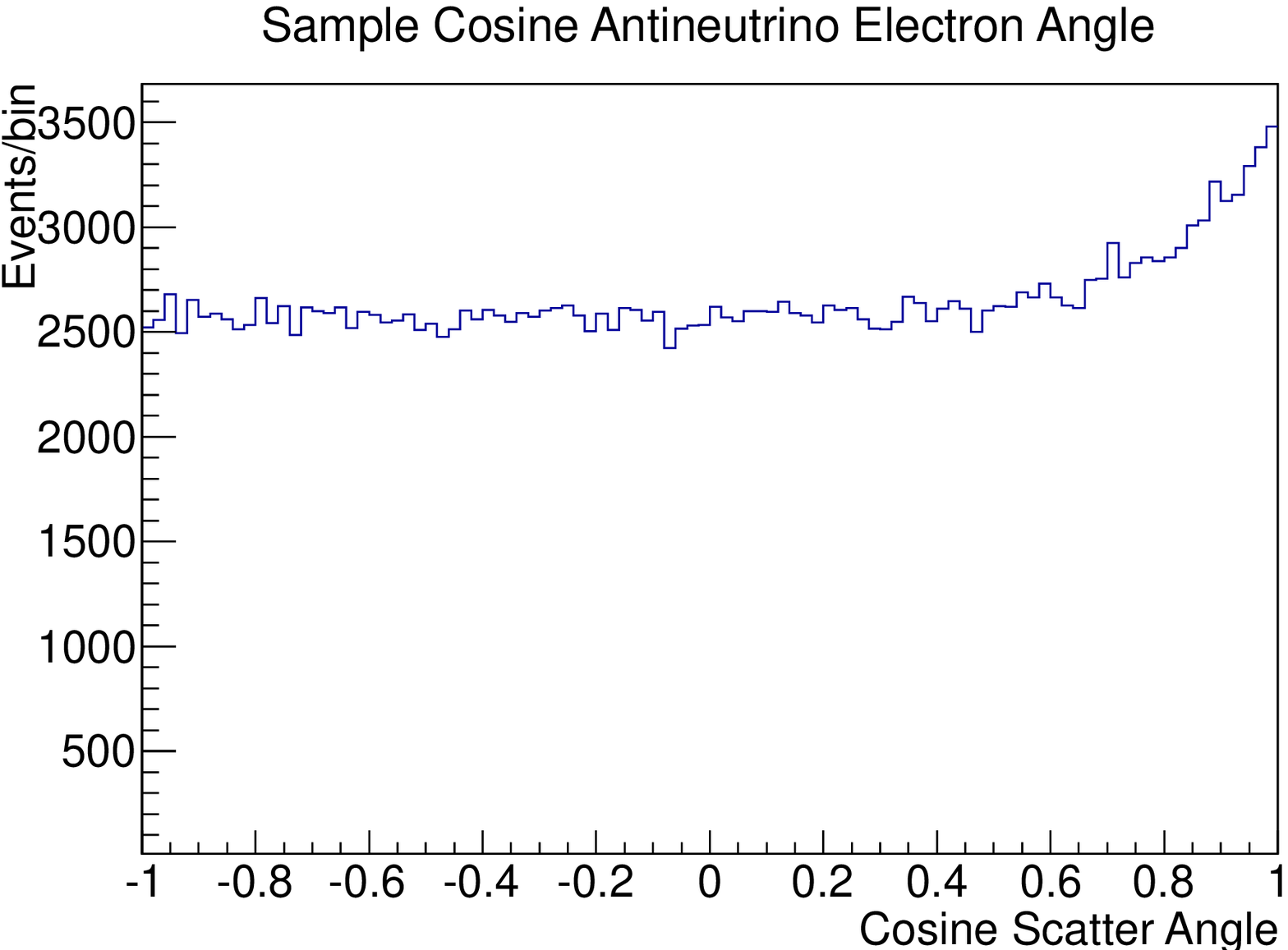}
\caption{The predicted water target detector response as a function of detected photoelectrons, according to the GEANT4 WATCHMAN simulation (left). The cosine of the angle between the reconstructed electron direction and the antineutrino direction for the electron scattering signal and background at WATCHMAN.  The predicted number of IsoDAR events and backgrounds correspond to 5 years run time. Events at cosine(angle) = 1 are defined as pointing directly away from the IsoDAR beam source (as expected for a scattered electron with zero scattering angle). Background events were sampled from a flat cosine(angle) distribution. }
\label{fig:NSIbackgrounds}
\end{center}
\end{figure} 

Table~\ref{tab:NSIevents} shows the beam and non-beam background predicted for a water Cherenkov WATCHMAN. Note that the misidentified IBD rate is 80 times higher than for scintillator, since the neutron tagging efficiency drops from $99.75\%$ to $80\%$.  Figure~\ref{fig:NSIbackgrounds} (left) shows the detector background response in WATCHMAN. Figure~\ref{fig:NSIbackgrounds} (right) shows the results of a GEANT4 simulation of electron scattering events in WATCHMAN, assuming a run time of 5 years. The Cherenkov cone reveals the direction of the scattered electron, the scattered electrons point back to the IsoDAR source. Here we plot the signal and background events as a function of the cosine of the scattering angle, which is approximately flat for background, and assumed to be flat here. We anticipate that the water Cherenkov WATCHMAN experiment will employ a statistical subtraction, based on scattering angle, to extract the spectral shape and integral number of electron scattering events.  

\begin{figure}
\begin{center}
\includegraphics[width=0.45\textwidth]{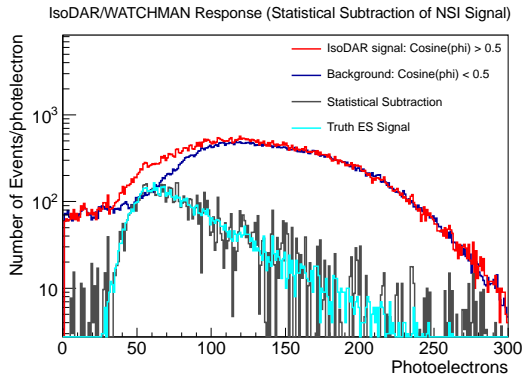}
\caption{The WATCHMAN detector response as a function of detected photoelectrons.  Shown here are the total background spectrum, the true scattered electron spectrum from the IsoDAR beam, the sum of the two (the detected WATCHMAN spectrum), and finally the background subtracted spectrum - the statistical subtraction of the backgrounds from the detected spectrum, based on the reconstructed electron-antineutrino scattering angle.}
 \label{fig:NSI_statisticalSubtraction}
\end{center}
\end{figure}

Figure~\ref{fig:NSI_statisticalSubtraction} shows the predicted visible spectrum from the electron scattering IsoDAR signal events together with all the backgrounds considered here. A statistical subtraction of the background was performed to extract the visible energy (in photoelectrons) of the elastic scattering signal using the scattering angle distribution.  

\subsubsection{Conclusions}

The IsoDAR NSI experiment will be a significant challenge at either KamLAND or WATCHMAN. For a liquid scintillator target, the WATCHMAN signal to background will be approximately $\sim 1:2.4$. At KamLAND it will be $\sim 1:1.4$.  The difference is primarily due to WATCHMAN's shallower depth, but is mitigated by the larger number of detected events. If high quantum efficiency PMTs are used at WATCHMAN the energy resolution will be better, potentially allowing for some further reduction in background and/or an enhanced energy spectrum. As it stands, we ignore the potential but uncertain gains from energy resolution and potential problems associated with not having a non scintillating buffer region between the target and the PMTs. To compare the statistical uncertainty of both the KamLAND and WATCHMAN scintillator options we must assume some beam off time is devoted to measuring the background. We assume that one year of beam off time can be obtained in both experiments.  The radionuclide background component however, can be measured while the beam is on, by analyzing the event rates immediately following muons. We assume that both KamLAND and WATCHMAN will be able to measure the rate of mis-ID IBD events to a vanishingly small uncertainty using calibration sources. With these assumptions the statistical significances predicted for KamLAND and WATCHMAN are $19.5 \sigma$ and $31.7 \sigma$ respectively after 5 years beam on and one year beam off. These results correspond to uncertainties of $5.1 \%$ and $3.2 \%$ respectively.

For the $Gd-H_{2}0$ target, to determine the statistical significance of the elastic scattering signal we fit a function consisting of the sum of a constant and exponential to the cosine scattering angle distribution.   

\begin{equation}  
 \label{nsi}
y = A + B e^{5.37x}
 \end{equation}

The slope of the exponential component was fixed at the value determined earlier (5.37) from a simulation of a large number of signal events - we assume that in WATCHMAN a directional calibration source will be available to help determine the exact shape of the signal component for the IsoDAR NSI experiment. The uncertainty of the integral number of directional elastic scattered electrons from IsoDAR therefore is given by the fractional uncertainty of the normalizing intercept of the exponential at $cosine (\theta) = +1$, which is $\pm 3\%$.  The final predicted number of NSI elastic scattering events from IsoDAR over five years for the $Gd-H_{2}O$ target is therefore $9010 \pm 270$, an uncertainty of $\pm 3.0 \%$.  The reason the statistical uncertainty of the $Gd-H_{2}O$ result is so small, even though the signal to background is small (relative to the scintillator option), is because scattering angle sensitivity allows for a more precise measurement of the background even while the beam is on.

\section{Conclusions}

WATCHMAN offers a compelling, cost-effective approach to exploiting the unique synergy between strategic research goals of the U.S. nonproliferation and physics communities.  The detector  will provide immediate physics benefit through its  tagging ability for supernova antineutrinos, and consequent sensitivity to the highly directional supernova neutrino signal. In a follow-on phase, it will provide a target for the ISODAR neutrino beam, enabling both a stringent sterile neutrino search and a search for non-standard neutrino interactions. Perhaps most importantly for the long-term future of U.S. neutrino physics, WATCHMAN provides a suitably large platform that is uniquely suited to testing of essential elements of very large neutrino detectors. The combination of Water Based Liquid Scintillator with Large Area Picosecond  Photodetectors in WATCHMAN will ensure the viability of these technologies for all follow-on detectors. 
In recognition of this overlap with the future U.S. physics program, the WATCHMAN collaboration has grown to include many of the country's most knowledgeable experts in water-based neutrino detection,  as well as members of the Advanced Scintillator Detector Concept, an initiative that is planning for the construction of a multi-hundred-ton WBLS detector in the United States. 
The benefit of WATCHMAN to the nonproliferation community is similarly important. WATCHMAN is a stepping-stone to true long range discovery or exclusion of small reactors. It provides a path to international, collaborative efforts at guaranteeing the absence of operating reactors in wide geographical regions, with important, positive implications for the nonproliferation regime.

\bibliographystyle{ieeetr}

\bibliography{WATCHMANwhitepaper}

\end{document}